\documentclass{aa}  

\usepackage{graphicx}
\usepackage{txfonts}
\usepackage{upgreek}
\usepackage{ulem}
\usepackage{arydshln}
\usepackage{multicol}
\usepackage{placeins}

\begin{document} 

    \title{The dark fate of ultra-faint dwarfs: Gravothermal collapse in action
    }
    \titlerunning{UFDs and the gravothermal collapse}

   \author{Moritz S.\ Fischer
          \inst{\ref{inst:dipc},\ref{inst:usm}}
          and
          Hai-Bo Yu\inst{\ref{inst:ucr}}
          }
    \authorrunning{M.\ S.\ Fischer and H.-B.\ Yu}

    \institute{
        Donostia International Physics Center (DIPC), Paseo Manuel de Lardizabal 4, 20018 Donostia-San Sebastián, Spain\label{inst:dipc}\\
        \email{moritz.fischer@dipc.org}
        \and
        Universitäts-Sternwarte, Fakultät für Physik, Ludwig-Maximilians-Universität München, Scheinerstr.\ 1, D-81679 München, Germany\label{inst:usm}
        \and
        Department of Physics and Astronomy, University of California, Riverside, California 92521, USA\label{inst:ucr}
    }

   \date{Received 6 March, 2026 / Accepted 26 May, 2026}

  \abstract{Ultra-faint dwarf (UFD) galaxies are a promising probe for dark matter (DM) physics as they are the most DM-dominated systems known. The Milky Way (MW) hosts many UFDs for which the properties of their DM distribution have been inferred from measurements of their stellar kinematics. If DM has self-interactions beyond gravity, the UFD halos may undergo a gravothermal evolution, giving rise to a population of galaxies with more diverse DM density profiles.}
  {We investigate DM densities of MW UFDs in self-interacting dark matter (SIDM) models, with an aim of determining the stage of gravothermal evolution for their halos.}
  {Therefore, we employed idealised high-resolution SIDM $N$-body simulations targeted to a MW-like system and compared the properties of simulated satellites to those of the observed UFDs.}
  {We find that the gravothermal evolution of SIDM halos produces diverse DM distributions, aligning with observations of the MW UFDs. Most of the UFDs have high DM densities, indicating that their halos have passed the period of maximum core expansion and entered the collapse phase, i.e.\ their central density may increase with time. The depth to which they have evolved into the gravothermal collapse may vary strongly across the satellites.
  This allows SIDM to account for the diversity in their DM densities. Moreover, the acceleration of the gravothermal evolution by tidal stripping can help to explain the diversity of the UFDs, as the ones with smaller pericentre distances require having evolved further into the gravothermal catastrophe.}
  {Large SIDM cross-sections of $\sigma / m_\chi \approx 80 \, \mathrm{cm}^2 \, \mathrm{g}^{-1}$ at a velocity of $v \approx 20 \, \mathrm{km} \, \mathrm{s}^{-1}$ are plausible, as the halo densities of MW UFDs are consistent with the gravothermal evolution predicted in SIDM, with most of them being in the collapse phase.}

   \maketitle

\section{Introduction}
\label{sec:introduction}

The cosmological standard model, Lambda cold dark matter ($\Lambda$CDM), has been remarkably successful in explaining the large-scale structure of the Universe and the overall properties of galaxies~\citep[e.g.][]{PlanckCollaboration_2020, Springel_2006, Trujillo-Gomez_2010}. On smaller scales, particularly in the central regions of galaxies, the success of collisionless CDM is less clear \citep[e.g.][]{Bullock_2017, Sales_2022}. Nevertheless, inferring the particle-physics properties of dark matter (DM) beyond its gravitational interactions from these small-scale observations remains challenging. This uncertainty arises both from observational limitations and from the difficulty of accurately modelling the complex baryonic physics that shapes galaxy formation. 

For example, in $\Lambda$CDM, the density of DM halos typically follows a Navarro-Frenk-White (NFW) \citep[]{Navarro_1996} profile with an inner logarithmic density slope of $-1$, which means a cusp. However, observations of many dwarf galaxies indicate shallow, cored density profiles \citep[e.g.][]{Oman_2015, Ren_2018}, motivating alternative models, such as self-interacting DM (SIDM)~\citep{Spergel_2000, Tulin_2018, Adhikari_2025}. Even within CDM, several astrophysical mechanisms can generate cores, including outflows driven by supernovae or active galactic nuclei \citep[e.g.][]{Governato_2012, Pontzen_2012, Martizzi_2013}, dynamical heating from massive perturbers or bars \citep[e.g.][]{ElZant_2001, Weinberg_2002}, and tidal stripping of satellite halos with an initially large radial anisotropy \citep{Chiang_2025}. Consequently, it is essential to disentangle these baryonic and dynamical effects from signatures of new DM physics \citep[e.g.][]{Burger_2022, Bosch_2026}.

Given these challenges, DM-dominated galaxies provide a valuable probe. Specifically, below a stellar mass threshold of $\approx 10^6 \, \mathrm{M_\odot}$, the supernova feedback becomes negligible for the formation of shallow density cores \citep[e.g.][]{Penarrubia_2012, Tollet_2016}. As their dynamics are mainly driven by the DM, this allows us to make a more unambiguous inference about the effect of new DM physics on the stellar distribution compared to more massive galaxies. Ultra-faint dwarf (UFD) galaxies can be viewed as a subclass of classical dwarf galaxies ($M_\star \approx 10^{5\text{--}7} \mathrm{M_\odot}$) and they are defined to have a luminosity of $L_\star \leq 10^5 \, \mathrm{L_\odot}$ and a stellar mass of $M_\star \approx 10^{2\text{--}5} \, \mathrm{M_\odot}$ \citep[e.g.][]{Bullock_2017, Simon_2019}.
Today, more than 60 UFD candidates associated with the Milky Way (MW) are known \citep{Richstein_2024, Pace_2025}, of which not all have been spectroscopically confirmed yet. These systems have been used to study various DM models, for example, primordial black holes \cite[e.g.][]{Graham_2024, Koulen_2026}, fuzzy DM \citep{Hayashi_2021b}, non-gravitational long-range interactions between DM and standard model particles \citep{Bogorad_2025}, or DM annihilation \cite[e.g.][]{Fisher_2025}.

Most MW UFDs appear broadly consistent with the cuspy density profiles predicted by CDM, although some may host central DM cores \citep[e.g.][]{Hayashi_2023, Errani_2024}. Notable candidates include Tucana~III~\citep{{Simon_2017}}, which reaches a small pericentric distance from the MW, and Eridanus~II~\citep{Li_2017}, whose off-centre star cluster is difficult to explain without invoking a central DM core \citep{Amorisco_2017, Simon_2021, Orkney_2022}. At the same time, some systems, such as Draco~II~\citep{Longeard_2018} and Phoenix~II~\citep{Fritz_2019}, may exhibit central densities even higher than CDM expectations, given current upper limits. Future observations may therefore reveal a subset of UFDs that are incompatible with collisionless DM.

In this work, we investigate DM distributions of UFDs in the SIDM framework, where collisions among DM particles locally thermalise the inner halo and alter its density. SIDM was originally introduced as a mechanism to reduce the abundance of satellite galaxies and to form density cores in DM halos \citep{Spergel_2000}. More recently, the influence of SIDM on various systems and measures has been studied \citep[see][for reviews]{Tulin_2018, Adhikari_2025}. An intriguing consequence of SIDM for the evolution of a DM halo is that it not only enables the formation of a low-density core, but it can also lead to gravothermal collapse, producing a high central density \citep[e.g.][]{Balberg_2002, Essig_2019, Feng_2020, Gad-Nasr_2024, Gurian_2025, Kamionkowski_2026, Fischer_2025, Feng_2025, Jiang_2025, Gu_2026, Li_2026}. In addition, the central density gradient remains flat during gravothermal collapse. 

This mechanism is promising for explaining several observational hints of highly concentrated DM halos \citep[e.g.][]{Minor_2020, Yang_2021, Nadler_2023, Silverman_2023, Dutra_2024, Enzi_2024, Kong_2025c, Roberts_2025, Tajalli_2025, Li_2025, Kong_2025a, Yu_2026, Kollmann_2025, Vegetti_2026}. For example, it has been investigated in the context of the perturber of the stellar stream GD-1 in the MW \citep{Zhang_2025, Fischer_2025}, which appears surprisingly compact and dense \citep{Bonaca_2019, Nibauer_2025}. Importantly, self-interactions that are sufficiently strong at low velocities to drive dwarf galaxies into the collapse phase emerge as a viable mechanism for diversifying the dwarf galaxy population \citep[e.g.][]{Zavala_2019, Kahlhoefer_2019, Sameie_2020, Nishikawa_2020, Turner_2020, Correa_2020, Correa_2022, Gilman_2022, Yang_2023, Fischer_2024a, Nadler_2023, Shah_2023, Nadler_2025b, Despali_2026, Jia_2026, Engelhardt_2026}.

In this paper, we evaluate a scenario in which most UFD galaxies associated with the MW, driven by DM self-interactions, have entered the gravothermal collapse phase. Our analysis is based on estimates of the dynamical mass within the half-light radius of UFDs, which we compare with high-resolution SIDM $N$-body simulations. We focus on UFDs with available spectroscopic data, enabling us to infer the line-of-sight velocity distribution and estimate their masses. We show that the gravothermal evolution of SIDM halos produces diverse DM distributions that align with observations of MW UFDs. Most UFDs exhibit high DM densities, indicating that their halos have passed the period of maximum core expansion and entered the collapse phase. The extent to which individual systems have progressed into gravothermal collapse varies across satellites, allowing SIDM to account for the observed diversity in their DM densities. We also demonstrate that UFDs with smaller pericentre distances tend to have smaller half-light radii, higher densities, and to be more evolved into the collapse phase.

The remainder of this paper is structured as follows. In Sect.~\ref{sec:simulations}, we describe the simulations that we use and discuss how representative their initial conditions (ICs) are. Subsequently, we compare the properties of the simulated satellite to the observed UFDs from the MW in Sect.~\ref{sec:results}.
In Sect.~\ref{sec:correlations}, we investigate correlations between the properties of the UFDs and their pericentre distance.
With Sect.~\ref{sec:discussion}, it follows a discussion on
possibilities for future improvements.
Finally, we summarise and conclude in Sect.~\ref{sec:conclusion}.
Additional information on correlations with the host distance of the UFDs is provided in Appendix~\ref{sec:appendix}.

\section{The simulations and their initial conditions}
\label{sec:simulations}

\begin{figure}
    \centering
    \includegraphics[width=\columnwidth]{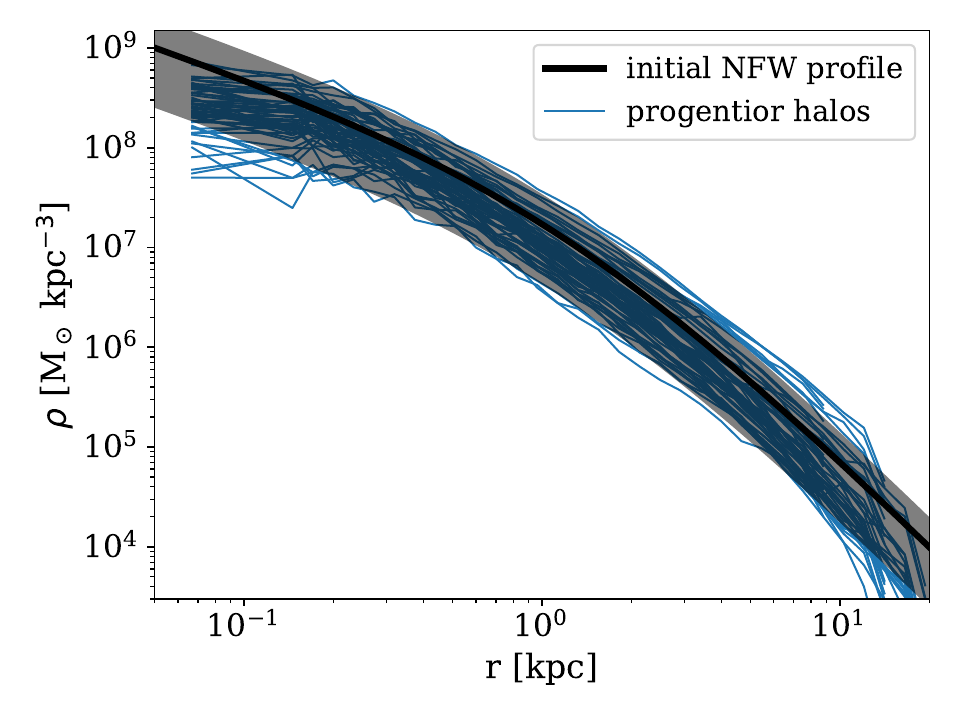}
    \caption{Initial NFW density profile \citep[as in][]{Fischer_2025} used for our simulations (black) and progenitor halos from a cosmological zoom-in simulation (blue). The latter are 60 halos from \cite{Yang_2023} within a mass range of $[5.6 \times 10^8, 5.6 \times 10^{9}] \, \mathrm{M_\odot}$, a subset of the halos shown by \cite{Zhang_2025}. The grey band indicates a density range between being a factor of five times lower and two times higher relative to the black curve. This range covers roughly the scatter in the cosmological zoom-in simulations and is used later in the paper.}
    \label{fig:progenitor}
\end{figure}

In this section, we first describe the simulations that we are using. Afterwards, we provide a motivation for the chosen initial conditions.

For our study, we employed the idealised CDM and SIDM simulations by \cite{Fischer_2025}, which were motivated to model the dense perturber of the GD-1 stellar stream \citep{Bonaca_2019, Zhang_2025}. In particular, we use their simulations T, W, and Y.
They were carried out with the $N$-body code \textsc{OpenGadget3} (Dolag et al.\ in prep.), a successor of the \textsc{Gadget-2} code \citep{Springel_2005}. It contains an SIDM module introduced by \cite{Fischer_2021a, Fischer_2021b, Fischer_2022, Fischer_2024a, Fischer_2026}.
The initial conditions follow an NFW profile \citep{Navarro_1996},
\begin{equation}
    \rho(r) = \frac{\rho_0}{\frac{r}{r_\mathrm{s}}\left(1+\frac{r}{r_\mathrm{s}}\right)^2} \,,
\end{equation}
with the density parameter being set to $\rho_0 = 4.42 \times 10^{7} \, \mathrm{M_\odot} \, \mathrm{kpc}^{-3}$ and the scale radius to $r_\mathrm{s} = 1.28 \, \mathrm{kpc}$. This corresponds to a halo with a mass of about $\approx 2.8 \times 10^9 \, \mathrm{M_\odot}$ at infall.

Moreover, the DM halo evolves within an external potential intended to mimic the influence of the MW. We refer the reader to Sect.~4.1 of~\cite{Fischer_2025} for a detailed description of the potential, which was motivated by the MW models of~\cite{McMillan_2011, McMillan_2016}, and summarise its key assumptions and caveats here. The potential is modelled analytically rather than resolved with simulation particles. It consists of six components: an NFW halo describing the MW's DM distribution, a stellar bulge modelled with a Hernquist profile profile~\citep{Hernquist_1990}, and thin and thick stellar and gas discs represented by axisymmetric Miyamoto–Nagai potential profiles~\citep{Miyamoto_1975}. We further assume that the potential is static in time, neglecting the evolutionary growth of the MW.

For the adopted NFW profile, the host halo has a virial mass $M_\mathrm{vir}\approx1.3\times10^{12}\,{\rm M_\odot}$ and a concentration $c\approx11.5$. We note that DM self-interactions can modify the inner structure of the MW halo, leading to deviations from an NFW profile. However, because the central MW is baryon dominated, SIDM thermalisation with the stellar potential produces a relatively small core and high central density~\citep{Kaplinghat_2013}, making the NFW approximation reasonable \citep[see Fig.~6 of][]{Sameie_2018}. In addition, we neglect interactions between SIDM particles belonging to the satellite and host halos. We discuss the justification for this assumption below.

The satellite is on an elliptical orbit with a pericentre distance of $d_\mathrm{peri} \approx 18 \, \mathrm{kpc}$ and an apocentre distance of $d_\mathrm{apo} \approx 142 \, \mathrm{kpc}$.
These orbital parameters are the same as those in \cite{Zhang_2025}, which were originally motivated to model the perturber of the GD-1 stellar stream.
Due to tidal stripping, the halo loses mass. At the end of the simulated time span, it has a mass of $\approx 2 \times 10^8\, \mathrm{M_\odot}$.
For further information on the simulations, see Sect.~2 by \cite{Fischer_2025}.

We want to highlight that the initial conditions of those simulations are representative of a DM halo falling into a MW mass-like halo. In Fig.~\ref{fig:progenitor}, we compare the initial density profile of our progenitor halo to those from the cosmological zoom-in simulations of an MW mass-like system by \cite{Yang_2023}. From the latter, we show 60 progenitor halos with a gravitationally bound mass in the range of $[5.6 \times 10^8, 5.6 \times 10^{9}] \, \mathrm{M_\odot}$ at infall, with data compiled in \cite{Zhang_2025}.
For the MW satellite galaxies, the masses of their progenitor halos are expected within this mass range.

\begin{figure*}
    \centering
    \includegraphics[width=\linewidth]{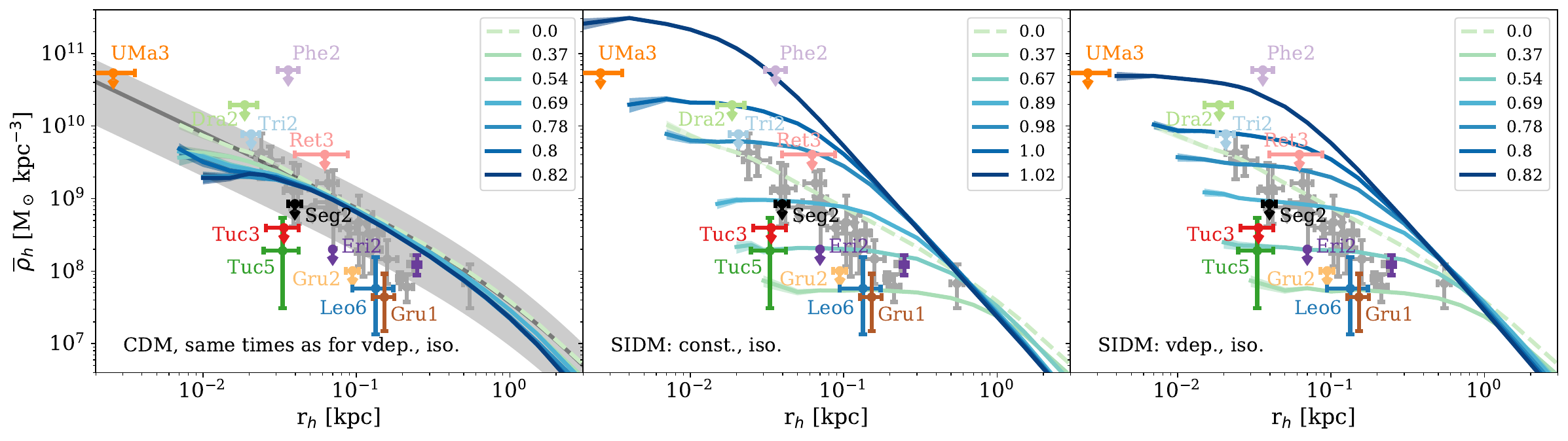}
    \caption{Average density as a function of radius. Left: collisionless run \citep[simulation T by][]{Fischer_2025}. Middle: velocity-independent run \citep[simulation W by][]{Fischer_2025}. Right: velocity-dependent run \citep[simulation Y by][]{Fischer_2025}. In the middle and right panels, the legend specifies the time relative to the collapse time ($\tau = t / t_*$), as specified by Eq.~\eqref{eq:collapse_time}. Whereas in the left panel, we show the results for the collisionless simulation for the same times as in the right panel. The dashed line for $\tau = 0$ corresponds to the initial conditions. Additionally, we display the data for various observed dwarf satellite galaxies as explained in the main text. To highlight some of them, we show them in colour and label them with their names. We note that the upper limit for Eridanus~II is not a measure of the density within the physical half-light radius but the density core inferred from its star cluster \citep{Orkney_2022}. Moreover, Ursa Major~III might be a star cluster instead of a dwarf galaxy \citep{Rostami_Shirazi_2025, Cerny_2026a}, but its nature remains ambiguous \citep{Adams_2026}. In the left panel, we additionally display a grey band to give an idea of the variation between CDM satellites. It is the same as in Fig.~\ref{fig:progenitor}, with the only difference that we are showing the average density here.
    The dark grey line is the analytic description of the initial NFW profile.
    }
    \label{fig:density_profile}
\end{figure*}

We consider three different DM models. They are collisionless DM (simulation T), SIDM with isotropic velocity-independent scattering (simulation W), and SIDM with isotropic velocity-dependent scattering (simulation Y). The total cross-section for the velocity-independent case is $\sigma/m_\chi = 80 \, \mathrm{cm}^2 \, \mathrm{g}^{-1}$. In the velocity-dependent case, the total cross-section is given by
\begin{equation}
    \frac{\sigma(v)}{m_\chi} = \frac{\sigma_0}{m_\chi} \left[ 1 + \left( \frac{v}{w} \right)^2 \right]^{-2} \, ,
\end{equation}
with $\sigma_0 / m_\chi = 6593.89 \, \mathrm{cm}^2 \, \mathrm{g}^{-1}$ and $w = 20 \, \mathrm{km} \, \mathrm{s}^{-1}$. As discussed previously, we model the MW potential analytically and neglect interactions between satellite and host DM particles. This approximation is justified provided that the self-scattering cross section on MW scales ($v\sim200\,\mathrm{km\,s^{-1}}$) remains below $\sim1\,\mathrm{cm^2\,g^{-1}}$. Accordingly, the constant cross section adopted in simulation W should be interpreted as an effective value appropriate for the satellite halo under consideration ($v\sim20\,\mathrm{km\,s^{-1}}$), arising from an underlying velocity-dependent SIDM model. For simulation Y, the cross section at $v\sim200\,\mathrm{km\,s^{-1}}$ is $0.6\,\mathrm{cm^2\,g^{-1}}$, and thus the above approximation is self-consistent.

The simulations are DM-only, in other words, no gas or stars are modelled. For the case of UFD galaxies, this should be appropriate as they contain very few stars and little gas and thus should not be significantly affected by them, as we discussed in Sect.~\ref{sec:introduction}.
Importantly, we want to point out that the satellite halos are resolved by $N = 10^7$ simulation particles. This allows us to model the inner $10^{-2}$--$10^{-1}$ kpc. This is something that is hardly possible in cosmological simulations as they require resolving the host system as well.
In addition, with the simulation scheme adopted in \cite{Fischer_2025}, the energy conservation in the SIDM collapse regime is under control, which can be challenging in simulations depending on the numerical scheme employed \citep[e.g.][]{Zhong_2023, Fischer_2024b, Mace_2024, Palubski_2024}.

Overall, we simulate only one set-up with different cross-sections. While the satellites of the MW evolve on different orbits, the gravothermal evolution within the long-mean-free-path regime is known to be universal \citep[e.g.][]{Lynden-Bell_1980, Balberg_2002} and also the evolution in the short-mean-free-path (SMFP) regime can be described in a universal manner to a good approximation \citep{Outmezguine_2023, Zhong_2023, Yang_2024, Gad-Nasr_2024}.
The tidal interactions break this strictly universal behaviour. However, they do not affect the qualitative evolution of the inner region of the halo much, but mainly alter the time evolution. Tidal stripping can speed up the evolution while tidal heating temporarily slows it down \citep[e.g.][]{Nishikawa_2020, Sameie_2020, Zeng_2022}. As a consequence, the evolution of our single set-up can be considered representative for most UFDs of the MW.

Moreover, the details of the cross-section do not alter the qualitative picture in the regime where the scattering of satellite particles with the host system's DM can be neglected. The strength of the cross-section controls the time-scale of the evolution, but the gravothermal evolution process of forming a density core and its subsequent collapse remains the same. Hence, the results that we find considering only two cross-sections are not only consistent with each other (see Sect.~\ref{sec:results}), but are also representative for many other models.

\section{Comparison to observed UFD satellites} \label{sec:results}

In this section, we compare the simulations by \cite{Fischer_2025} to observations of UFD satellite galaxies. 
In particular, we study their half-light radii and the average density within these radii.

Here, we consider 35 satellites of the MW. They are mainly known as UFDs, but some of them could be star clusters, as we will discuss later. For most of them, we use the data compiled in the Local Volume Database \citep{Pace_2025}, where dynamical masses are based on the \cite{Wolf_2011} estimator.
In detail, we consider the following UFDs:
Aquarius~II \citep[Aqu2,][]{Bruce_2023},
Aquarius~III
\citep[Aqu3,][]{Cerny_2026a},
Boötes~I \citep[Boo1,][]{Sandford_2026},
Boötes~II \citep[Boo2,][]{Bruce_2023},
Boötes~III \citep[Boo3,][]{Carlin_2009},
Canes Venatici~II
\citep[CVn2,][]{Simon_2007},
Centaurus~I
\citep[Cen1,][]{Heiger_2024},
Columba~I \citep[Col1,][]{Fritz_2019},
Coma Berenices \citep[Com,][]{Simon_2007},
Draco~II \citep[Drac2,][]{Longeard_2018},
Eridanus~II \citep[Eri2,][]{Li_2017},
Eridanus~IV \citep[Eri4,][]{Heiger_2024},
Grus~I \citep[Gru1,][]{Chiti_2022},
Grus~II \citep[Gru2,][]{Simon_2020},
Hercules \citep[Her,][]{Simon_2007},
Hydra~II \citep[Hya2][]{Kirby_2015},
Leo~IV \citep[Leo4,][]{Jenkins_2021},
Leo~VI \citep[Leo6,][]{Tan_2025},
Pegasus~III \citep[Peg3,][]{Kim_2016},
Pegasus~IV \citep[Peg4,][]{Cerny_2023},
Phoenix~II \citep[Phe2,][]{Fritz_2019},
Pisces~II \citep[Psc2,][]{Kirby_2015},
Reticulum~III \citep[Ret3,][]{Fritz_2019},
Segue~I \citep[Seg1,][]{Simon_2011},
Segue~II \citep[Seg2,][]{Kirby_2013},
Triangulum~II \citep[Tri2,][]{Buttry_2022},
Tucana~II \citep[Tuc2,][]{Chiti_2021},
Tucana~III \citep[Tuc3,][]{Simon_2017},
Tucana~IV \citep[Tuc4,][]{Simon_2020},
Tucana~V \citep[Tuc5,][]{Hansen_2024},
Ursa Major~I \citep[UMa1,][]{Simon_2007},
Ursa Major~II \citep[UMa2,][]{Simon_2007},
Ursa Major~III \citep[UMa3,][]{Cerny_2026a},
and Willman~I \citep[Wil1,][]{Willman_2011}. 
We note that Ursa Major~III is now considered more a star cluster than a UFD \citep{Rostami_Shirazi_2025, Cerny_2026a}. But we give it as a reference point nonetheless. For Eridanus~II, we additionally consider the density core inferred from its off-centre star cluster \citep{Orkney_2022}.

\begin{figure}
    \centering
    \includegraphics[width=\columnwidth]{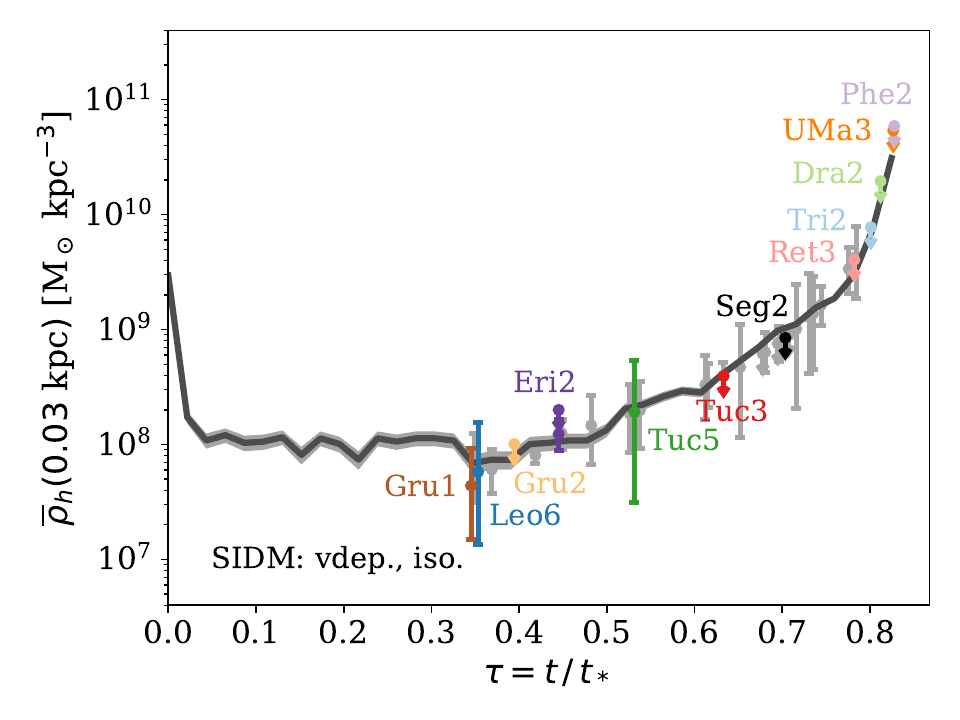}
    \caption{Central density as a function of time.
    The velocity-dependent run (simulation Y by \cite{Fischer_2025}) is used to show the average central density of the halo within 0.03 kpc as a function of time relative to the collapse time (Eq.~\eqref{eq:collapse_time}). In addition, the same UFD satellite galaxies as in Fig.~\ref{fig:density_profile} are given.
   They are assigned $\tau$ values chosen to approximately match the densities from the simulations; these $\tau$ values should not be overinterpreted as precise measurements.
    We show the results for the velocity-dependent cross-section only, as the results for the velocity-independent cross-section \citep[simulation W by][]{Fischer_2025} are qualitatively very similar.
    }
    \label{fig:central_density}
\end{figure}

In practice, we identify the snapshot at which the density of the simulated halo is aligned with the measured one of a UFD. This approach is justified with the following considerations. Although the initial conditions are not tailored to model individual UFDs in general, the gravothermal evolution of SIDM halos is approximately self-similar, as discussed in the previous section, and hence the matched density from the simulation can be representative. Moreover, for a given SIDM model, there are degeneracies among initial halo properties and tidal orbital parameters in determining the stage of the gravothermal evolution. A UFD can have an orbital trajectory different from the one in the simulation, but we can still get a very similar SIDM density profile by shifting the halo parameters, particularly the concentration.

For comparison, we make use of the SIDM collapse time as the normalisation factor to characterise the evolution time, motivated by \cite{Zhong_2023} and \cite{Yang_2024}, where they showed in terms of a normalisation factor that the evolution of the SIDM density profile can be expressed universally.
For a halo initially following an NFW profile, it can be parametrised as \citep{Balberg_2002, Koda_2011, Essig_2019}
\begin{equation} \label{eq:collapse_time}
t_* = \frac{150}{\beta} \left(\frac{\sigma}{m_\chi}\right)^{-1} \frac{1}{\rho_\mathrm{0}\,r_\mathrm{s} \sqrt{4 \,\uppi \, \mathrm{G}\, \rho_\mathrm{0}}} \, .
\end{equation}
Given that all halos were evolved with an effective cross-section of $80 \, \mathrm{cm}^2 \, \mathrm{g}^{-1}$ \citep{Yang_2023}, the corresponding collapse time according to Eq.~\eqref{eq:collapse_time} is $4.51 \, \mathrm{Gyr}$, where we have employed a value of $0.704$ for $\beta$ following the $\beta_\mathrm{eff}$-model by \cite{Mace_2026}.
In the following, we will give the time in units of the collapse time $t_*$, in other words, we specify the time as $\tau = t / t_*$.
We note that for the halo evolving in the tidal field, the core-collapse time changes due to mass loss and tidal heating. In principle, such effects could be modelled \citep[see, e.g.][]{Yang_2024}. In this work, we use Eq.~\ref{eq:collapse_time} for simplicity, and leave a detailed investigation for future work.

\subsection{Density profile}

We begin the comparison by studying the average three-dimensional matter density $\overline{\rho}_\mathrm{h}$ within the radius $r_\mathrm{h}$ and compare it to the data of observed dwarf satellites discussed above. Figure~\ref{fig:density_profile} gives the corresponding density profile for the simulations assuming a collisionless model (T, left panel), a constant cross-section (W, middle panel), and a velocity-dependent cross-section (Y, right panel) by \cite{Fischer_2025}. In each panel, the density profiles are specified by the ratio of the evolution time to the collapse time (Eq.~\eqref{eq:collapse_time}).
For the UFDs, the density was computed based on the estimated dynamical mass within their half-light radius. 

The left panel shows the comparison of the density profiles in the CDM simulation and the averaged densities within the half-light radii. There is an overall trend of the observed UFDs that $\rho_h$ increases as $r_h$ decreases, aligned with that of the CDM halos. In fact, most of the UFDs are consistent with the CDM halo profiles after taking into account the spread or variation. Nevertheless, UFDs also demonstrate a great diversity in the density, as Tucana~III, Eridanus~II, and Grus~II have too low densities, while others may feature too high densities, such as Draco~II and Phoenix~II, though those are upper limits, and the actual DM contents could be lower. Importantly, to explain those with low densities in CDM, their progenitor halo masses would be close to $5.0 \times 10^8 \mathrm{M_\odot}$, about one order of magnitude higher than the star formation threshold \citep{Nadler_2025a}.
Overall, the majority of UFD satellites of the MW can be well explained in CDM, and only a few appear to be outliers.

In the middle and right panels of Fig.~\ref{fig:density_profile}, one can see how the density core forms starting from the initial NFW profile indicated by the dashed line. As we are interested in the collapse phase, the solid lines are picked from the phase where the central density is increasing as a function of time.
If all the SIDM halos are in the core-expansion phase, this means close to the density profile with $\tau = 0.37$ shown in the middle and right panels of Fig.~\ref{fig:density_profile}, the majority of UFDs cannot be explained.
For those with low densities, such as Tucana~III, Eridanus~II, and Grus~II, their halos are likely to be in the core-expansion phase or shortly thereafter.
Although the simulated density profiles are from different snapshots, meaning different $t$, of the single SIDM halo, the results are valid in general.
Here $\tau$ should be interpreted as an indicator of the relative phase of gravothermal evolution.
Turning this around, suppose that we assume $t$ to be, for example, $10$ Gyr for all UFDs, we would need to adjust $t_*$ to keep individual $\tau$ values fixed.
This means that for low-density galaxies, the $\tau$ values are large, indicating a long collapse time.
For a given cross-section, it means that the halo concentration $c$ is low as $t_* \propto c^{-7/2}$ \citep{Essig_2019,Nadler_2023}, while for those with high densities, the concentration is high, resulting in a small $t_*$ and a large value for $\tau$. As a consequence, SIDM cannot only explain the UFDs, which are consistent with CDM, but also the ones that appear to be outliers in a collisionless DM model; in other words, SIDM accounts for the diversity of the observed systems.

To further distinguish SIDM predictions from CDM, it would be promising to investigate UFDs with potentially very high central DM densities, such as Draco~II, Phoenix~II, and Ursa Major~III.
From current measurements, there are only upper limits on the densities within $r_h$ for the first two systems (as indicated in Fig.~\ref{fig:central_density}), while the latter one may not be a UFD.
If future measurements of Draco~II and Phoenix~II show that their densities are indeed high and close to their upper limits, this would be a strong indication of gravothermal collapse induced by DM self-interactions. 

Interestingly, Fig.~\ref{fig:density_profile} shows that it is even more challenging to distinguish between SIDM simulations W and Y using the UFD measurements. This is because, by construction, the simulated halos follow similar gravothermal evolution trajectories in the two SIDM models, as they yield similar effective cross sections governing the evolution of the halo under consideration (Appendix~C of~\cite{Fischer_2025}). Distinguishing between the two scenarios would therefore require additional observations of galaxies on larger mass scales. Nevertheless, an important conclusion we can draw is that SIDM models with a cross section of $\mathcal{O}(1)\,\mathrm{cm^2\,g^{-1}}$ on UFD scales are disfavoured, as in such models most halos would remain in the core-expansion phase by a Hubble time, leading to central densities too low to be consistent with the measurements of most UFDs.

Accurately estimating the dynamical mass of UFDs can be challenging. Especially, when they contain only a few stars for which the line-of-sight velocities are measured to estimate their mass. Obtaining the line-of-sight velocity distribution is complicated by the presence of binary stars. If unidentified, they may cause an overestimate of the velocity dispersion of the system. This is because of their increased velocity when orbiting each other compared to a single stellar system that is only affected by the global potential of the dwarf galaxy. As a consequence, the dynamical masses for some UFDs could be overestimated \citep[e.g.][]{Pianta_2022}. With the help of multiepoch observations, binary systems can be identified and the binary pollution largely avoided. For example, for Ursa Major~III, \cite{Cerny_2026a} obtained spectroscopic data for more member stars and identified further binary systems compared to previous work \citep[e.g.][]{Errani_2024}. They suggest that the system is more likely a star cluster than a UFD \citep[see also][]{Devlin_2025, Rostami_Shirazi_2025}.
A more general study modelling UFDs with DM-free stellar systems was undertaken by \cite{Flammini_Dotti_2026}.
For systems that contain sufficiently many stars, one can statistically model their effect. This allows to reduce the problem of binary pollution and is in particular helpful if there is only a single epoch of observations available \citep[e.g.][]{Gration_2025}.

While a star cluster could be confused for a dwarf galaxy, there is also the possibility that a UFD is not hosted inside a DM halo but in the gravitational well of a black hole. This possibility has been modelled for Segue~I by \cite{Lujan_2025}, one of the UFDs that we consider. They pointed out that there is no need for a spatially extended DM halo, but a black hole is sufficient to explain the observations.

\subsection{Central density}
Next, we study the temporal evolution of the central density.
In Fig.~\ref{fig:central_density}, we show the average density within $r=0.03 \, \mathrm{kpc}$ of the simulated halos as a function of time. In addition, we display the densities for the UFD sample. However, for those, we do not directly give the average density within $r=0.03 \, \mathrm{kpc}$, but the inferred values for their densities with their half-light radii.
For most systems, it appears reasonable to compare those values. This is because for time instances of our simulated halos that could possibly explain the UFDs the density gradient within their half-light radius is typically flat such that computing the average density within their half-light radius from the simulations would not make much of a difference.
To demonstrate that the observed UFDs could be explained with various stages of the gravothermal evolution of the simulated satellite halo, we assigned them a time $\tau$ such that they match the simulated central density. These values are meant only as a rough indication of how far the UFDs could have evolved into gravothermal collapse. A more precise determination is complicated by the interplay of tidal forces and self-interactions, which act on different time scales and can give rise to a divers phenomenology \citep[e.g.][]{Klemmer_2026}. Similarly, \cite{Kong_2025a, Roberts_2025} previously used $\tau$ to characterise the gravothermal evolution phase of isolated field galaxies, whereas we apply it to satellite galaxies.

The core formation phase, where the central density drops can be seen well in Fig.~\ref{fig:central_density}. It extends to almost $\tau = 0.4$. Subsequently, the collapse phase follows, where the central density increases.
It is visible that most UFDs of our sample plausibly could be in the collapse phase.
For those systems where comparing the average density within $r=0.03 \, \mathrm{kpc}$ to the one within the half-light radius works the least well, we underestimate a little how far the systems would have evolved into the gravothermal collapse. So some of them could plausibly have even higher values of $\tau$.

\subsection{Gravitationally bound mass}

\begin{figure}
    \centering
    \includegraphics[width=\columnwidth]{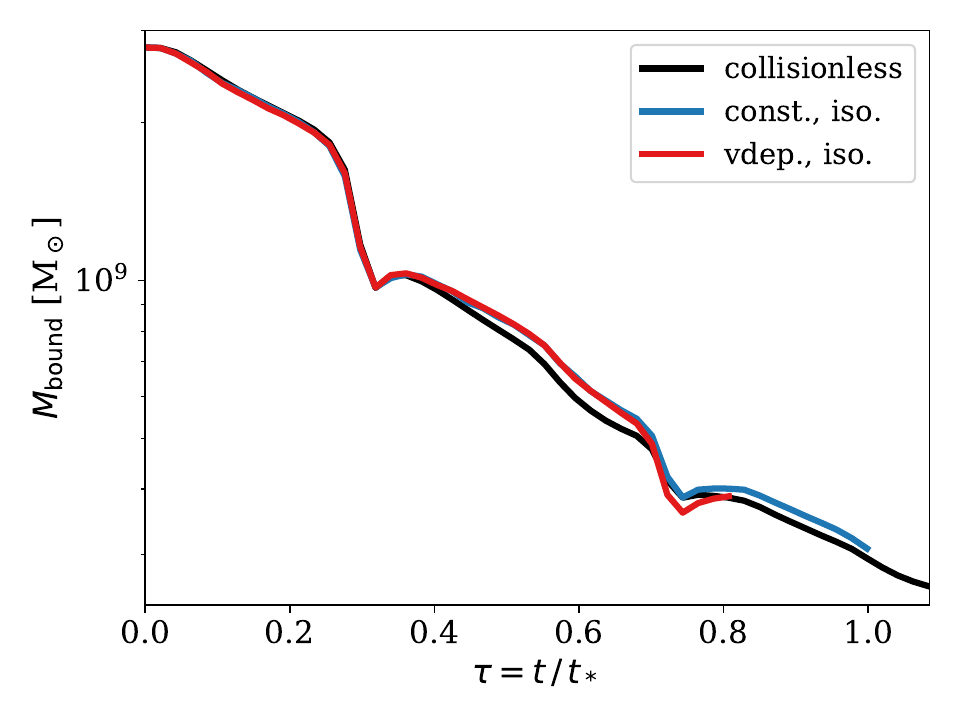}
    \caption{Gravitational bound mass as a function of time. The bound mass for the different DM models, as indicated in the legend, is shown. The mass was computed by considering particles that initially belonged to the subhalo only and ignoring the host system.}
    \label{fig:bound_mass2}
\end{figure}

The satellite halos lose mass due to tidal interaction with the host galaxy. While the sample of observed UFDs that we are using may span a range of gravitationally bound mass. The simulated system evolves in mass and thus also covers a corresponding range, making it more representative of the UFD sample to which we compare it.

In Fig.~\ref{fig:bound_mass2}, we show the gravitationally bound mass as a function of time.\footnote{When evaluating the bound mass, we neglect the gravitational potential of the host.} It is clearly visible that the mass decreases over time. The local minima in the bound mass correspond roughly to the pericentre passages, where tidal forces are strongest. Overall, the mass of the simulated system decreases by roughly one order of magnitude over the period we study. We note that the mass loss is similar between the different DM models considered here.
The mass reached by the end of the simulated period is reasonable for hosting the MW's UFDs.

\section{Correlations} \label{sec:correlations}

In this section, we show how properties of the MW's UFDs vary with the distance of their orbital pericentre to the MW. \cite{Kaplinghat_2019} pointed out an anti-correlation between the central DM density of bright MW dwarf spheroidal galaxies and their orbital pericentre. This behaviour may provide a promising probe for SIDM models \citep[e.g.][]{Kahlhoefer_2019, Sameie_2020, Ebisu_2022, Hayashi_2023}.
Similarly, we study the average density within the physical half-light radius, $r_h$. For SIDM halos, the density of DM within $r_h$ is almost constant, and thus the average density is representative of the central density of the DM halo (see also Fig.~\ref{fig:central_density}).
As a consequence, comparing $\overline{\rho}_h$ for the UFDs is meaningful in the context of DM self-interactions and may tell us something about their gravothermal evolution. For the pericentre distance, we rely mainly on the values given by \cite{Pace_2022}, for a few systems not given there, we use the ones provided by \cite{Simon_2018}.

\begin{figure}
    \centering
    \includegraphics[width=\columnwidth]{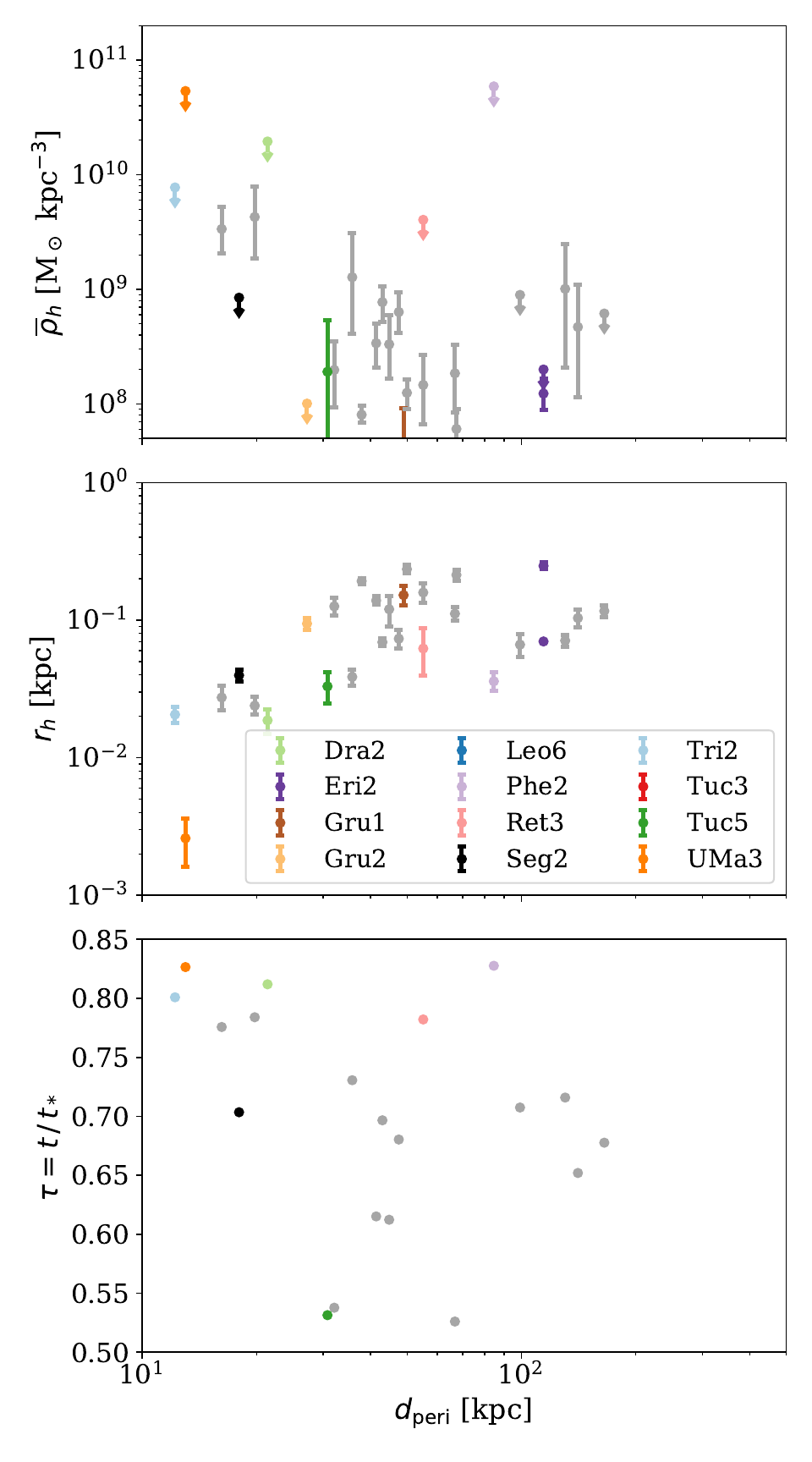}
    \caption{Correlations of the MW UFDs with the pericentre distance of their orbit. Upper panel: The average density within the physical half-light radius is displayed as a function of the pericentre to host distance. Middle panel: The physical half-light radius of the UFDs is shown. Lower panel: Stage of gravothermal evolution as a function of the pericentre distance. We use the gravothermal evolution stage that we assigned to the UFDs in Fig.~\ref{fig:central_density}.}
    \label{fig:correlations_peri}
\end{figure}

The upper panel of Fig.~\ref{fig:correlations_peri} shows the average density within the half-light radius versus the pericentre distance for the MW UFDs, revealing an overall strong anti-correlation. In the SIDM framework, a natural expectation is that UFDs with smaller pericentre distances experience stronger tidal stripping, which accelerates gravothermal collapse and drives the halo into a deeper collapse phase with higher central density~\citep[e.g.][]{Nishikawa_2020, Sameie_2018}. In addition, a smaller pericentre may also indicate earlier halo infall and thus a higher halo concentration, further promoting deeper collapse and enhanced density.

We note that the anti-correlation for the UFDs appears stronger than reported in previous studies~\citep{Kaplinghat_2019, Hayashi_2023}. \cite{Kaplinghat_2019} first showed that such a trend is pronounced for the brightest MW dwarfs, but becomes weak for the UFD population. \cite{Hayashi_2023} further argued that the scatter is large and found no clear evidence for an anti-correlation among UFDs. One important difference is that these previous works characterised the density using the inferred value at $150\,\mathrm{pc}$ obtained from fitting a specific density profile, whereas we instead use the average density within the half-light radius. We leave it to future work to investigate whether this difference systematically affects the assessment of the strength of the anti-correlation.

The middle panel reveals that the physical half-light radius tends to decrease for small pericentre distances. It is expected that the tidal interactions with the MW strip matter from the satellites and thus can potential lead to smaller half-light radii of the UFDs. Tidal forces are stronger for satellites close to the host. As UFDs with small pericentre distances come closer to the MW than those with a larger pericentre distance, they experience stronger tidal interactions, potentially contributing to the observed trend of the half-light radius to decrease with pericentre distance.
In particular, more concentrated halos are less affected by tidal forces, potentially giving rise to a `survivor bias' \citep[e.g.][]{Hayashi_2020, Genina_2022}.

In the lower panel, we show the gravothermal evolution time, $\tau$, that we assigned based on the average density in Fig.~\ref{fig:central_density}. It illustrates that within an SIDM framework, one can interpret the halos with higher densities to have evolved further into the gravothermal collapse. These satellites that might have evolved further are, on average, closer to the host where the tidal forces are stronger. Those forces are known to accelerate the gravothermal evolution by tidal stripping \citep[e.g.][]{Nishikawa_2020,Sameie_2020} and thus qualitatively align with the shown results.
However, we must note as a caveat that the large scatter among the UFDs makes it difficult to draw a robust conclusion.

\section{Discussion} \label{sec:discussion}

In this section, we comment on the findings that can be drawn from comparing theoretical predictions to the observed UFDs and discuss other relevant studies. Moreover, we elaborate on the limitations of our work and highlight directions to improve upon them. This concerns aspects regarding the observations of UFDs as well as modelling their evolution. 

We compare our analysis with previous constraints on SIDM using MW UFDs. Neglecting the gravothermal collapse scenario, \cite{Hayashi_2021a} derived a tight constraint on the cross-section, $\sigma/m_\chi < \mathcal{O}(0.1)\,\mathrm{cm}^2\,\mathrm{g}^{-1}$, using the kinematics of the UFDs Segue~I and Willman~I. When gravothermal collapse is included, \cite{Ando_2025} employed a semi-analytical method and found the cross-section at low velocities to be $\sigma/m_\chi \leq 1\,\mathrm{cm}^2\,\mathrm{g}^{-1}$, along with a preferred range at much larger cross-sections, $\sigma/m_\chi \gtrsim 10^{2\text{--}3}\,\mathrm{cm}^2\,\mathrm{g}^{-1}$. The latter is consistent with our results indicating that UFDs could predominantly be in the collapse phase. \cite{SanchezAlmeida_2025} constrained the cross-section to $0.35\,\mathrm{cm}^2\,\mathrm{g}^{-1} < \sigma/m_\chi < 199\,\mathrm{cm}^2\,\mathrm{g}^{-1}$ for UFDs, based on the assumption that a core in the stellar distribution reflects a DM density core; this constraint also allows for the possibility of core collapse. Despite differences in the inferred cross-section among these studies, in the SIDM framework most observed MW UFDs are consistent with being in the collapse phase.

Concerning the comparison of simulated satellites to observed ones, a single system that is consistent with CDM may always appear to be consistent with a specific state of the gravothermal collapse phase when comparing only the average density within the physical half-light radius. This makes it difficult to distinguish a CDM halo from a core-collapsed SIDM halo. However, if a statistically sufficient sample of dwarf galaxies is known, it becomes possible, as the gravothermal evolution implies how many systems one would expect to be in a specific state, or in other words, what the distribution of $\tau$-values would be.

To make inferences about the collapse phase, we need a better understanding of the observed UFDs and their DM distributions beyond the average density, although challenges remain in practice. For example, many UFDs show a density core in their stellar distribution, which could eventually reflect an underlying DM distribution that is cored as well \citep{SanchezAlmeida_2024}. Nevertheless, a cuspy DM halo could also host a stellar distribution with a shallow density gradient \citep{Haekkinen_2026}. \cite{Errani_2023} pointed out that systems on orbits with a small pericentre cannot have a large DM core, as tidal forces may disrupt the satellites otherwise, and this in particular concerns Tucana~III, Segue~I, Segue~II, Reticulum~II, Triangulum~II, and Willman~I. However, this may not apply to core-collapsed SIDM halos, as they are more resilient to core-expansion halos~\citep[]{Kong_2025b}, and dedicated SIDM simulations tailored to these UFDs can help further improve the assessment. Additionally, it would be fruitful to extend the SIDM analysis to the population of ultra-faint compact satellites, which are smaller in size compared to the UFDs~\citep{Cerny_2026b}. 

There are ways to improve on the modelling of UFDs. One of them is to include the scattering between the DM particles of the satellite and the host. This has been explored by \cite{Kummer_2017, Zeng_2022, Klemmer_2026} when using an analytic description for the host halo. Moreover, the inclusion of collisionless particles to mimic the evolution of the stellar component would allow for a better comparison with observations. The tagging method to assign DM particles some stellar mass, as employed by \cite{Zhang_2024}, may not be applicable to the collapse phase. However, it could be applied based on a dynamically negligible collisionless component, and thus would allow us to study the evolution of stellar components with various initial density profiles without running a simulation for every stellar distribution.

The simulations we use assume an initially isotropic DM velocity distribution, whereas a more realistic setup would allow for anisotropy. In the collisionless case, anisotropy can significantly affect satellite evolution by enhancing mass loss; in particular, stripping from highly radial orbits can produce central density cores \citep{Chiang_2025}. In SIDM, \cite{Kamionkowski_2025} showed for isolated systems that velocity anisotropy impacts the collapse time. More generally, anisotropy effectively modifies the heat conductivity in the gravothermal fluid model \citep{Bettwieser_1986}. In addition, we have considered only a single halo on a specific orbit. In reality, UFDs arise from a diverse population of halos with varying masses, concentrations, stellar properties, and orbital parameters. Modelling this full distribution would enable a more precise comparison with observations.

UFDs provide a promising avenue to constrain a broader class of SIDM models. In two-species scenarios with inter-species interactions, the central density can follow a different evolutionary path \citep{Patil_2025, Yang_2025}; in particular, core collapse may be delayed or even halted, thereby avoiding the gravothermal catastrophe that is unavoidable in the one-species elastic SIDM model considered here. If a substantial population of UFDs exhibits large $\tau$ values but only moderately enhanced central densities, rather than the extreme densities predicted by our fiducial model, this would favour the two-species scenario. Conversely, if DM self-interactions include dissipative channels, energy loss can accelerate core collapse \citep{Essig_2019, Huo_2019}, producing central densities well above CDM expectations on timescales much shorter than the Hubble time. UFDs can therefore place upper limits on the allowed energy loss in DM self-interactions.

\section{Conclusion} \label{sec:conclusion}

In this work, we compared a sample of UFDs associated with the MW to CDM and SIDM $N$-body simulations of a satellite halo. In particular, we employed dynamical mass estimates and studied the average density within the physical half-light radius. Moreover, we evaluated at which stage of the SIDM induced gravothermal evolution the satellite galaxies could be. Gravothermal evolution of SIDM halos produces diverse DM distributions, aligning with observations of the MW UFDs. Most of the UFDs have high DM densities, indicating that their halos have entered the collapse phase. The depth to which they have evolved into the gravothermal collapse across the satellites, producing the diversity in the UFD densities. we also showed that that UFDs with smaller pericentre distances tend to have smaller half-light radii, higher densities, and to be more evolved into the collapse phase. Overall, UFDs are a promising probe for DM, and future observations are expected to detect more systems and reduce the error margins on their properties. This will not only enable us to shrink the parameter space of viable DM models, but also unambiguously allow detecting a DM signature beyond collisionless cold DM, eventually the gravothermal evolution driven by DM self-interactions.

\begin{acknowledgements}
It is with pleasure that we acknowledge the Local Volume Database\footnote{\url{https://github.com/apace7/local_volume_database}} \citep{Pace_2025} used for some of the dwarf galaxy data shown in the figures.
MSF gratefully acknowledges the support of the Alexander von Humboldt Foundation through a Feodor Lynen Research Fellowship.
HBY acknowledges support by the U.S.\ Department of Energy under grant No.\ de-sc0008541 and the John Templeton Foundation under Grant ID\# 61884.  The opinions expressed in this publication are those of the authors and do not necessarily reflect the views of the funding agencies.\\
Software:
NumPy \citep{NumPy},
Matplotlib \citep{Matplotlib}.
\end{acknowledgements}

\bibliographystyle{aa}
\bibliography{references.bib}

\begin{appendix}
\FloatBarrier
\section{Correlations with host distance} \label{sec:appendix}

\begin{figure}
    \centering
    \includegraphics[width=\columnwidth]{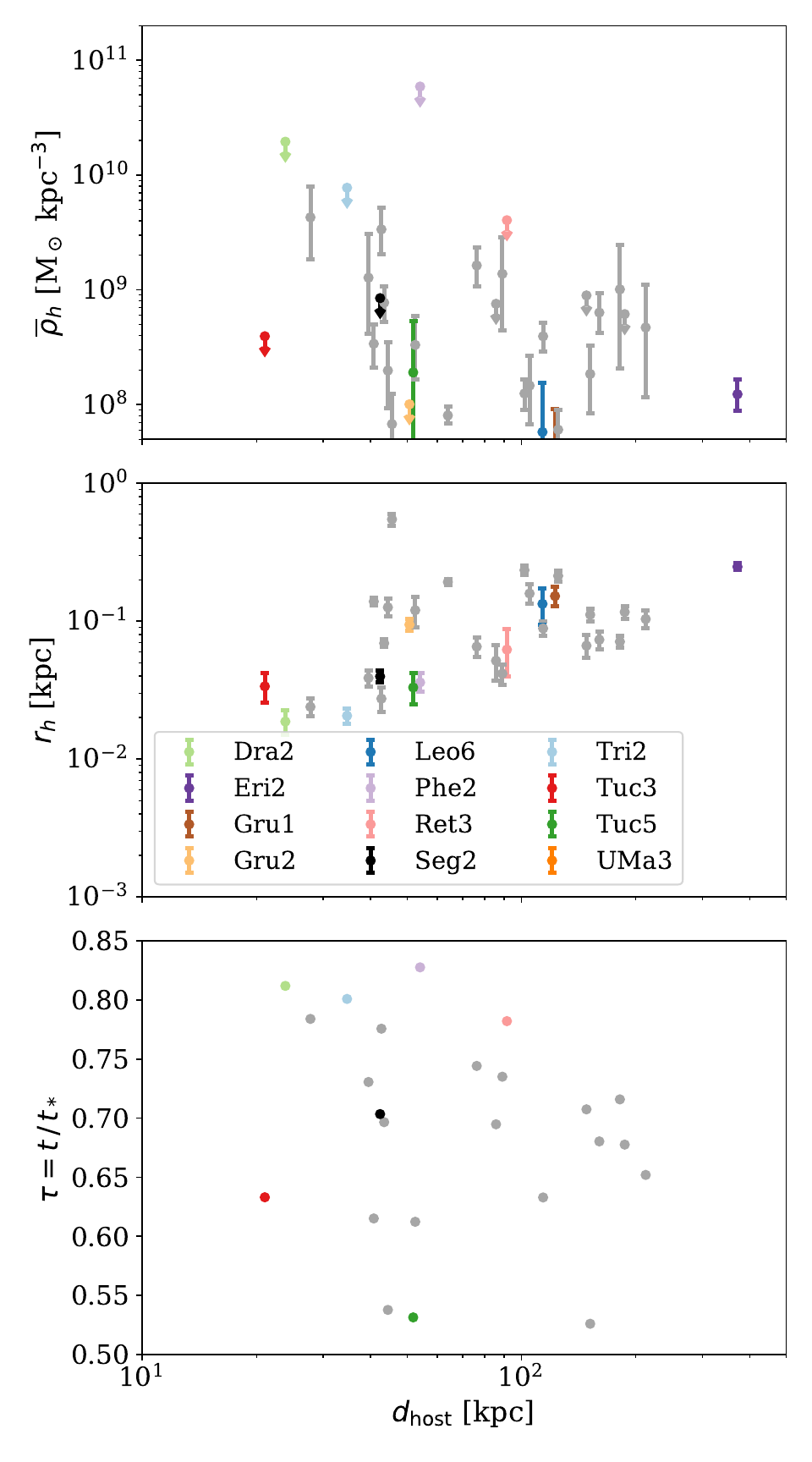}
    \caption{Correlations of the MW UFDs with the distance to the MW. The same as in Fig.~\ref{fig:correlations_peri}, but for the distance to the host, instead of the pericentre distance. The upper panel gives the average density within the physical half-light radius, the middle panel gives the physical half-light radius, and the bottom panel gives the gravothermal evolution time $\tau$.}
    \label{fig:correlations_dist}
\end{figure}

In Sect.~\ref{sec:correlations}, we showed the average density within the physical half-light radius, the physical half-light radius and the gravothermal evolution time $\tau$ as assigned in Fig.~\ref{eq:collapse_time} as a function of the pericentre distance for the UFDs of the MW.
Here we provide the same as in Fig.~\ref{fig:correlations_peri}, but for the distance between the centre of the MW and the current position of the UFDs. The distance of the satellites to the host is taken from the Local Volume Database \citep{Pace_2025}. In contrast to the pericentre distances, they do not depend on the modelling of the satellite's orbit.

The results are displayed in Fig.~\ref{fig:correlations_dist}. They reveal the same trends as in Fig.~\ref{fig:correlations_peri} for the pericentre distance.
In the upper panel, the average density within the physical half-light radius is given as a function of the satellite's distance to the MW. It is visible that the UFDs closer to the centre of the MW have higher densities, as one would expect for SIDM.
The middle panel gives the physical half-light radius as a function of the host distance. It tends to become smaller for systems closer to the host. This could be understood in terms of tidal force due to the host's gravitational potential removing stars from the satellite and leading to a smaller half-light radius.
The bottom panel gives the gravothermal evolution time, which directly reflects the higher densities of the systems closer to the MW and illustrates that this can be interpreted as these systems have evolved further gravothermally compared to systems at larger distances.

\end{appendix}

\end{document}